# Magnetoresistance Oscillations in Granular Superconducting Niobium Nitride Nanowires

U. Patel[1,2], Z. L. Xiao*[1,2], A. Gurevich[3], S. Avci[1], J. Hua[1,2], R. Divan[4],

U. Welp[2], and W. K. Kwok[2]

[1] Department of Physics, Northern Illinois University, DeKalb, Illinois 60115
[2] Materials Science Division, Argonne National Laboratory, Argonne, Illinois 60439
[3] National High Magnetic Field Laboratory, Tallahassee, Florida 32310
[4] Center for Nanoscale Materials, Argonne National Laboratory, Argonne, Illinois 60439

We report on magnetoresistance oscillations in superconducting $NbN_x$ nanowires synthesized through ammonia gas annealing of $NbSe_3$ precursor nanostructures. Even though the transverse dimensions of the nanowires are much larger than the superconducting coherence length, the voltage-current characteristics of these nanowires at low temperatures are reminiscent of one-dimensional superconductors where quantum phase slips are associated with the origin of dissipation. We show that both the magnetoresistance oscillations and voltage-current characteristics observed in this work result from the granular structure of our nanowires.

Superconducting nanowires have recently received intense attention[1-16]. On one hand, they are highly desirable in future electronic nanodevices. For example, nanowires of zero-resistance are ideal interconnects since they can circumvent the damaging heat produced by energy dissipation in a normal nano-conductor whose high resistance is inversely proportional to its cross-section area. Furthermore, in the resistive state they can act as superconducting quantum interference devices[9,10]. On the other hand, superconducting nanowires provide unique experimental testbeds to investigate and discover novel superconducting phenomena in confined geometries: Falk et al. probed dynamics of a few-row vortex lattice in $NbSe_2$ nanowires[15] and Tian et al. reported an anti-proximity effect in Zn nanowires with bulk superconducting electrodes[12]. Quasi one-dimensional (1D) superconducting nanowires with diameters comparable to the zero-temperature superconducting coherence length $\xi_0$ have been the research subject of thermal and quantum phase slip phenomena which induce dissipations at temperatures near and away from the superconducting critical temperature, respectively.

The observation of quantum phase slips (QPS) associated with the long resistance tail in the resistance versus temperature (*R-T*) curves at low temperatures is of significance not only for 1D superconductor but also for understanding the decoherence of a quantum system due to interaction with their environment[14]. However, interpretation of these results can be flawed by the presence of granularity in the nanowires that could give rise to a similar temperature dependence of the resistance[4,7,13,17]. Since it is extremely challenging in experiments to eliminate granularity, an alternative way to improve the understanding of the dissipation in 1D superconducting nanowires is to study superconducting granular nanowires in the absence of QPS and compare their properties with those observed in 1D nanowires.

In this Letter we report experiments on *free-standing* superconducting $NbN_x$ nanowires which are stable in ambient atmosphere, enabling the attachment of gold (Au) electrodes for standard four-probe measurements. We observed intriguing magnetoresistance oscillations, which we attribute to the granular nature of the nanowires. More importantly, in addition to the long tail in the *R-T* curves, we observed specific characteristics in the voltage-current (*V-I*) measurements which resemble those[2,13] reported in 1D superconductors where QPS were believed to be the origin for the low temperature dissipation. Since QPS should be absent in our $NbN_x$ nanowires whose transverse dimensions are much larger than the superconducting coherence length, $\xi_0$, and granularity does not depend on the dimensionality of the wire, our results indicate that the interpretations of previous QPS experimental results are inconclusive.

$NbN_x$ nanowires with critical temperatures up to 11 K and transverse dimensions down to tens of nanometers were synthesized by annealing $NbSe_3$ nanostructure precursors in flowing ammonia gas at temperatures up to 1000°C [Ref.18]. By utilizing standard photolithography and lift-off process, we deposited four gold electrodes with a gap of 5 μm between voltage leads onto individual nanowires using magnetron sputtering. We carried out angular dependent DC magneto-transport measurements in the constant current mode in a magnet-cryostat system which enabled precise stepper-motor controlled sample rotation in a magnetic field (with an angle resolution of 0.053°). Data reported here are from two $NbN_x$ wires with widths and thicknesses of $w = 350$ nm, $d = 160$ nm (sample A) and $w = 500$ nm, $d = 320$ nm (sample B), respectively. The zero-field critical temperature $T_{c0}$ defined with a 50% normal state resistance $R_N$ criterion is 9.94 K and 7.25 K for samples A and B, respectively.

The main panels and the upper-left insets of Fig.1 present the essential finding of this research. In the low-field regime of the magnetic field dependence of the resistance (*R-H*) curves we observe reproducible and pronounced oscillations in both samples. Similar magnetoresistance oscillations were observed in Nb-Ti/Cu multilayers in parallel magnetic fields and interpreted as the dynamic matching of a moving vortex lattice with the periodic microstructure[19]. However, we find significant differences between the data in multilayers and our current results: the magnetoresistance oscillations in multilayers persist up to the normal state while in our samples they occur only at low dissipation levels (less than 20%$R_N$) and in the low field regime where the change of the magnetoresistance is relatively weak. More importantly, the dynamic matching induced oscillations[19] disappear at low driving currents, in contrast to the data presented in Fig.2(a) which show that the amplitude of the oscillation decreases with increasing current.

Figure 2(b) presents the fast Fourier transform (FFT) power spectrum of the magnetoresistance of sample A at 9.2 K. The FFT spectrum indicates that the magnetoresistance oscillation is quasi-periodic. The dominant peak at $f = 55$ T$^{-1}$ in the FFT spectrum corresponds to a period of $\Delta H = 0.018$ T. Similarly we obtained a period of $\Delta H = 0.013$ T for the dominant oscillation in the $R$-$H$ curves of sample B.

The observed oscillation periods are consistent with two physically different field scales, which are of the same order of magnitude. The first one is the lower critical field $H_{c1}$ in a thin film strip of width $w$ smaller than the London penetration depth, $\lambda$. For NbN$_x$, $\lambda(0)$ at zero temperature is about 200 nm. Thus, for sample A ($T_{c0} = 9.94$K) at $T \geq 9$K, we have $\lambda(T) = \lambda(0)[T_{c0}/(T_{c0}-T)]^{1/2} \geq 650$ nm $> w$. In this case $H_{c1} \cong (2\Phi_0/\pi^2 w^2)\ln(w/\xi)$ where $\Phi_0$ is the flux quantum[20] sets the field increment required for penetration of the first few vortex rows resulting in resistance and magnetization oscillations[20]. For $w = 300$ nm and $w/\xi = 100$ characteristic of our nanowires, $H_{c1} \cong 20$ mT is consistent with the oscillation periods in Fig.2(b). Given the rough edges of the NbN nanowires, one can hardly expect an ideal penetration of parallel vortex rows. Instead, a more likely scenario is the penetration of mesoscopic vortex segments through different regions of suppressed edge barriers along the nanowire. The incoherent penetration of such vortex bundles can produce resistive oscillations characterized by a distribution of nearly temperature independent periods ~ $H_{c1}$, in agreement with our data. However, both theories[21,22] and experiments[21,23] demonstrate that the period of the oscillation originating from the vortex rearrangement should increase with increasing number of vortex rows. This differs from our data which show no change of the period with increasing field.

Another mechanism of oscillations can result from the granular structure of the nanowires which was proposed by Herzog et al. to explain the magnetoresistance oscillations observed in their *in-situ* grown granular Sn nanowires: screening currents circulating around phase coherent loops of weakly linked superconducting grains[5]. In this case, electrical transport is limited by critical currents $I_c$ across the grain boundaries (GB) which can be treated as short ($w<\lambda$) Josephson junctions. Such GB network can also produce the magnetotransport oscillations associated with the Fraunhoffer oscillations of $I_c(H)$ through a thin film Josephson junction of length $L$. These oscillations will have field period, $\Delta H \cong 1.8\Phi_0/L^2$ [Ref.24]. In a narrow nanowire, the distribution of GB lengths $L = w/\cos\alpha$ of straight GBs spanning across the entire cross section is determined by the distribution of local angles $\alpha$ between a GB and the edge of the nanowire. For $w = 300$ nm, we obtain $\Delta H_i = 40\cos^2\alpha_i$ mT, consistent with the magnitude and the temperature independence of the oscillation periods. Additional resistance harmonics with smaller field periods may result from spatial inhomogeneities along GBs, for example, faceting[25] and penetration of first vortex rows. As the field increases, further flux penetration in the grains washes out the Fraunhoffer oscillations, resulting in two distinct regimes in the *R-H* curve and apparent oscillations occur in the low-field weak magnetoresistance regime. It is also understandable that the nanowires are granular since the replacement of Se atoms by N atoms during the conversion from NbSe$_3$ to NbN$_x$ at high temperatures will cause atomic scale rearrangements, resulting in a change in the morphology of the nanowire. In fact, both AFM (see inset of Fig.2(a)) and SEM images indicate the granular morphology of the NbN$_x$ nanowires with an average grain size of 20-30 nm and also show the existence of tiny voids[18].

Experiments on the magnetic field orientation effect on the resistance oscillation also provide strong support for the above interpretation. The inset of Fig.3(a) presents *R-H* curves obtained with the magnetic field applied at angles from the perpendicular ($\phi = 0°$) to the parallel ($\phi = 90°$) long wire axis direction. Oscillations can be seen in all *R-H* curves and the period increases when the applied field is tilted towards the wire's long axis. The angular dependence of the oscillation period can be understood given the fact that only the perpendicular component of the magnetic field can penetrate GBs. Thus the period $\Delta H$ should increase as $1/\cos\phi$. Indeed, this dependence was revealed by the experimental data plotted in the main panel of Fig.3(a) for the dominant period. Considering that a loop of intragranular screening current may not be precisely contained in the *x-I* plane, the fitting and the experimental data are reasonably consistent. Data shown in Fig.3(b) provide further evidence: the same tendency is also observed for oscillations in the *R-H* curves with the magnetic field rotated in the plane perpendicular to the wire axis. In this case the oscillation period or the first peak field $H_1$ as defined in the inset, again follows the relation ~ $1/\cos\theta$ at angles up to 70°.

As mentioned above, there is a long running debate on the origin of the resistance tail in *R-T* curves of quasi 1D superconducting nanowires, since both QPS and granularity of the nanowires can induce similar effects. Both analysis of the nanowire morphology and magnetoresistance oscillations revealed that our NbN$_x$ nanowires are granular. The *R-T* curves presented in the lower-right insets of Fig.1 also indicate that their granularity (coupling between grains) is adjustable: the transition widths measured between 10% and 90% of $R_N$ are 0.66 K and 2.65 K for samples A and B, respectively. The normal state resistivity of sample B (10.6 mΩ·cm) is about 50 times larger than that of sample A (0.2 mΩ·cm). Inset of Fig.4(b) also shows that the *R-T* curve of sample B has a long resistance tail with an exponential decay similar to those observed in thin Sn nanowires[13]. Furthermore, our NbN$_x$ nanowires are not 1D. Thus they can serve as a model system for probing granular properties of superconducting nanowires in the absence of QPS, revealing benchmarks useful in understanding the low temperature dissipation in 1D superconducting nanowires.

Figures 4(a) and (b) show features in the *V-I* characteristics of NbN$_x$ nanowires similar to those[2,13] observed in quasi 1D superconducting nanowires where QPS is believed to contribute to the dissipation: the appearance of ohmic finite resistance at low currents and at temperatures far way from $T_c$ in sample B which has weak grain coupling, is similar to those found in smaller diameter (40 nm and 20 nm) Sn nanowires[13]. Although the voltage jumps and steps observed in the *V-I* curves of our samples resemble those found in small Sn nanowires, the similarity can be attributed to the self-heating effect which naturally occurs in granular nanowires. In fact, as presented in the inset of Fig.4(a) the current *I\** where the first voltage jump occurs follows the square root temperature dependence predicted by the self-heating model[26]. As shown by the *V-I* curve obtained at 2 K for sample A, similar voltage 'flip-flop' as that reported in Sn nanowires with diameter of 70 nm also occurs in our NbN$_x$ granular nanowires. This voltage 'flip-flop' phenomenon can be understood again with the grain loop model used to interpret the magnetoresistance oscillation: if the critical current densities ($J_c$) of the two current flowing branches of a loop are different, the branch with a lower $J_c$ will first become normal, resulting in a voltage jump which routes the current in this branch to the second branch which has a higher $J_c$; at a certain time during the current routing, both branches are superconducting before the second branch becomes normal while the first branch is superconducting. This current-routing process repeats itself, leading to a voltage oscillation with time.

Another interesting phenomenon related to the QPS experiments is the oscillatory structure in the *V-I* curves of 1D PbIn wires observed by Giordano et al. . Their explanation was based on the quantum tunneling rate change due to the current dependence of the quantum levels[2]. This oscillatory structure can even evolve into a voltage minimum, as reported by Tian et al. in their thinnest Sn nanowires[13]. As shown in Fig.4(b), similar structures including the voltage minimum also appear in *V-I* curves of our NbN$_x$ nanowires at low temperatures. In a granular nanowire, this can be explained by an increase in the quasiparticle tunneling rate due to the suppression of superconductivity in the grains arising from current induced depairing and/or self-heating[27].

In summary, we observed magnetoresistance oscillations in *free-standing* superconducting NbN$_x$ nanowires. Our experiments have shown that these oscillations can result from the Fraunhoffer oscillation of intergranular critical currents. We also found that the granularity of the nanowire can induce similar features in the voltage-current characteristics as those commonly attributed to quantum phase slips.

This material is based upon work supported by the US Department of Energy Grant No. DE-FG02-06ER46334 and Contract No. DE-AC02-06CH11357. S. A. was supported by the National Science Foundation (NSF) Grant No. DMR-0605748. The nanocontacting and morphological analysis were performed at Argonne's Center for Nanoscale Materials (CNM) and Electron Microscopy Center (EMC), respectively.

*Corresponding author, xiao@anl.gov or zxiao@niu.edu

**Figure captions**

**FIG.1** (Color online). *R-H* curves of samples A (*I*=2 μA) and B (*I*=0.5 μA). Upper-left insets show expanded views of the low-field data at numerous fixed temperatures. The magnetic field direction is perpendicular to the wire axis. Zero-field *R-T* curves are given in the lower-right insets.

**FIG.2** (Color online). (a) *R-H* curves of sample A at 9.2 K and various applied currents. Inset shows an AFM image of sample B. (b) FFT trace from $R_{osc}$-*H* curve presented in the inset for sample A at 9.2 K, where $R_{osc}$ is the oscillatory part of the *R-H* curve at *I* =2 μA.

**FIG.3** (Color online). Period of the dominant oscillation Δ*H* and the first peak field $H_1$ at various angles. The solid lines represent the inverse cosine relationship. The definitions of angles $\phi$ and $\theta$ are given in the lower-right insets. *R-H* curves at various angles are presented in the upper-left insets where the curves at angles other than 90° have been offset for clarity.

**FIG.4** (Color online). *V-I* characteristics at various temperatures in zero field. The inset in (a) gives the temperature dependence of the current *I*\* at which the first voltage jump occurs, where the solid symbols and line are experimental data and the fit, respectively. Inset of (b) presents the zero-field *R-T* curve of sample B with resistance in logarithmic scale.

**FIG.1**

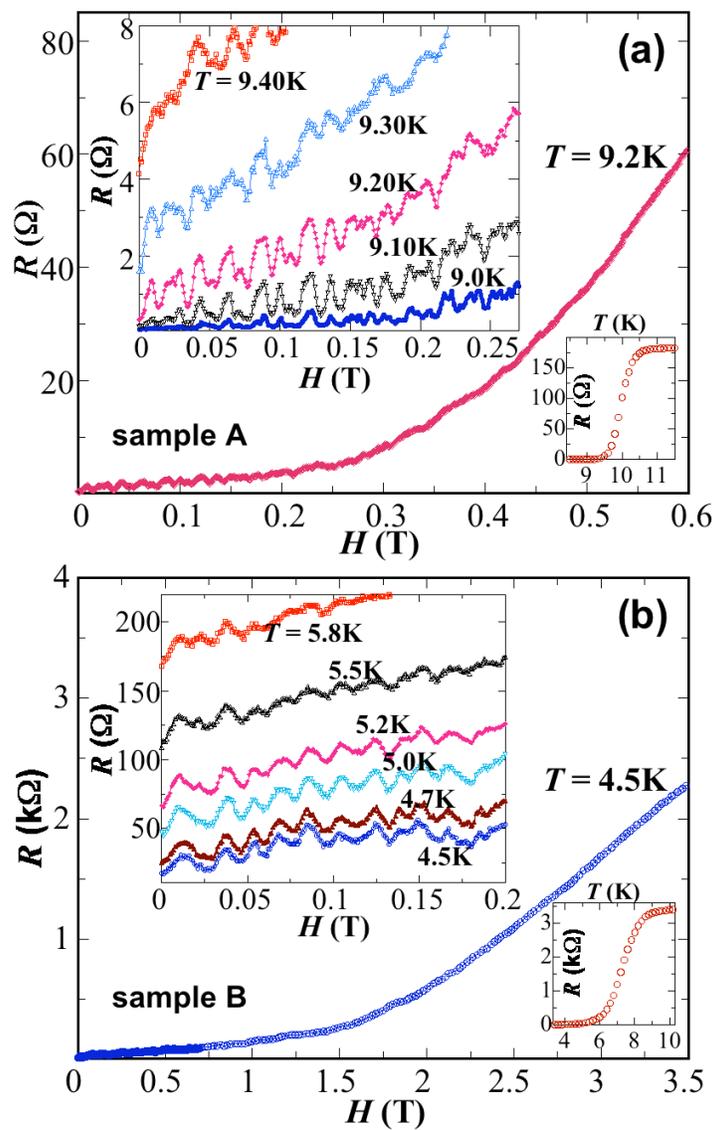

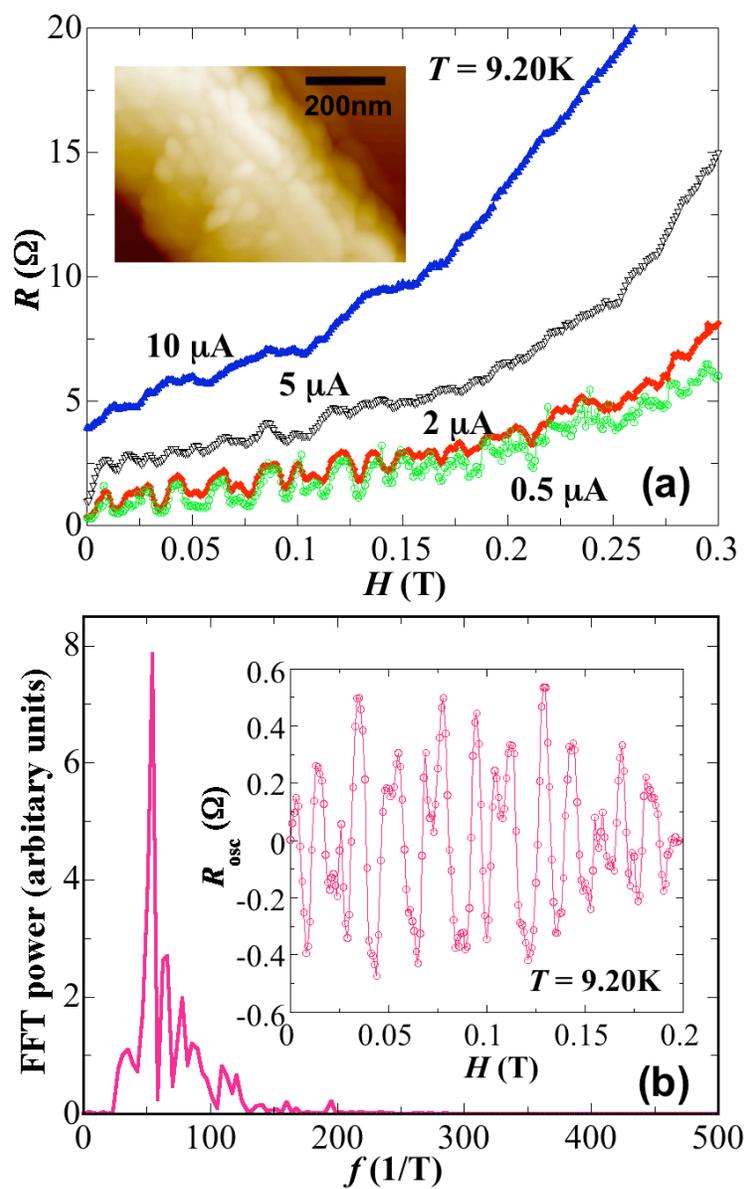

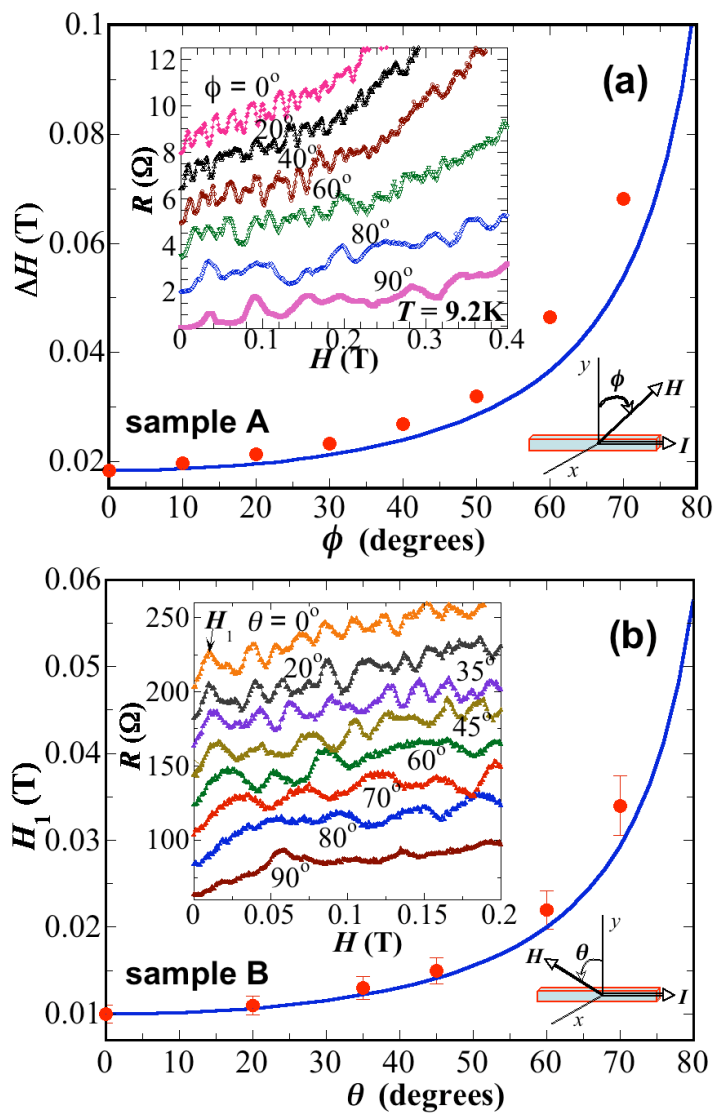

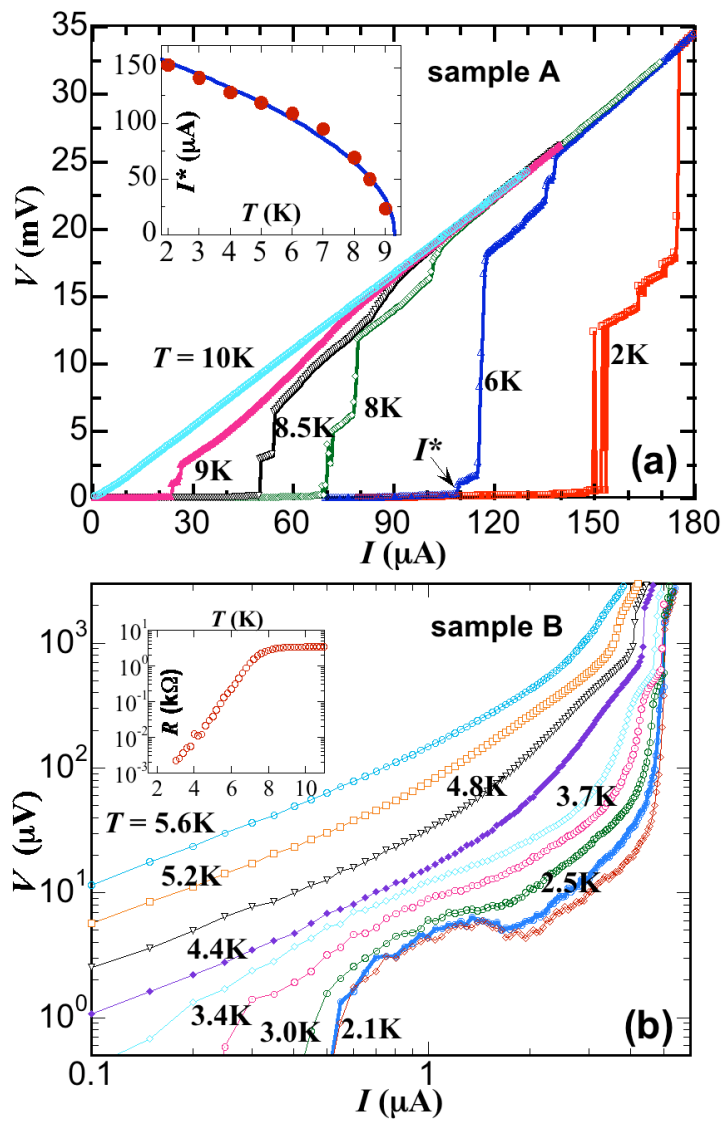